\documentclass[useAMS,usenatbib]{mn2e}
\usepackage{graphicx}                                            
\usepackage{times}
\usepackage{epsfig}
\usepackage{rotating,amssymb,xspace}

\def\lesssim{\mathrel{\hbox{\rlap{\hbox{\lower4pt\hbox{$\sim$}}}\hbox{$<$}}}}
\let\la=\lesssim
\def\gtrsim{\mathrel{\hbox{\rlap{\hbox{\lower4pt\hbox{$\sim$}}}\hbox{$>$}}}}
\let\ga=\gtrsim

\def\deg{\hbox{$^\circ$}}
\def\farcs{\hbox{$.\!\!^{\prime\prime}$}}

\def\lx{L_\mathrm{X}}

\def\Fx{$F_\mathrm{X}$}

\def\Nh{N$_{H}$}

\def\chired{$\chi^{2}_{\nu}$}
\def\chis{$\chi^{2}$\xspace}

\newcommand{\swiftj}{Swift~J1357.2--0933\xspace}

\def\simlt{\mathrel{\rlap{\lower 3pt\hbox{$\sim$}}
        \raise 2.0pt\hbox{$<$}}}

 \title[Multiwavelength spectral evolution of \swiftj]{Multiwavelength spectral evolution during the 2011 outburst of the very faint X-ray transient \swiftj}

   \author[M.~Armas Padilla et al.]{M.~Armas Padilla$^{1}$ \thanks{E-mail:M.Armaspadilla@uva.nl}, N.~Degenaar$^{2}$, D.~M.~Russell$^{3}$ and R.~Wijnands$^{1}$\\
$^{1}$Astronomical Institute "Anton Pannekoek", University of Amsterdam, Postbus 94249, 1090 GE Amsterdam, The Netherlands\\
$^{2}$University of Michigan, Department of Astronomy, 500 Church St, Ann Arbor, MI 48109, USA\\
$^{3}$Instituto de Astrof\'isica de Canarias (IAC), V\'ia L\'actea s/n, La Laguna 38205, S/C de Tenerife, Spain}
\begin{document}

\maketitle

\begin{abstract}

We report our multiwavelength study of the 2011 outburst evolution of the newly discovered black hole candidate X-ray binary \swiftj. We analysed the \textit{Swift} X-ray telescope and Ultraviolet/Optical telescope (UVOT) data taken during the $\sim$7~months duration of the outburst. It displayed a 2-10~keV X-ray peak luminosity of ~$\sim10^{35}$(D/1.5 kpc)$^{2}$~erg~s$^{-1}$ which classifies the source as a very faint X-ray transient. We found that the X-ray spectrum at the peak was consistent with the source being in the hard state, but it softened with decreasing luminosity, a common behaviour of black holes returning to quiescence from the hard state. The correlations between the simultaneous X-ray and ultraviolet/optical data suggest a system with a black hole accreting from a viscous disc, and we do not detect X-ray reprocessing on the disc surface. The UVOT filters provide the opportunity to study these correlations up to ultraviolet wavelengths, a regime so far unexplored. If the black hole nature is confirmed, \swiftj would be one of the very few established black hole very-faint X-ray transients.

\end{abstract}


\section{Introduction}

X-ray binaries are the brightest X-ray point sources in our Galaxy. They are black holes (BH) or neutron stars (NS) accreting material from a companion star. 
The transient X-ray binaries alternate long epochs of quiescence during which they have X-ray luminosities of $\lx\sim10^{30-33}$~erg~s$^{-1}$ with outburst episodes during which the luminosity increases more than two orders of magnitude. The systems that reach a peak X-ray luminosity in their outbursts of only $L^{peak}_{X}\sim10^{34-36}$~erg~s$^{-1}$  are called \textit{very-faint} X-ray binary transients (VFXTs). They are  several orders of magnitude fainter at their peaks than the better studied \textit{faint} ($L^{peak}_{X}\sim10^{36-37}$~erg~s$^{-1}$) and \textit{bright}  ($L^{peak}_{X}\sim10^{37-39}$~erg~s$^{-1}$) systems \citep{wijnands2006}.\\
It is in the last decade that VFXTs have been investigated in detail thanks to the improvement in sensitivity and resolution of the X-ray instruments in orbit. However, even though the number of known sources has increased, their characteristics are still not well known. A considerable fraction of them have exhibited thermonuclear Type-I X-ray bursts, identifying the accretor as a NS (eg. \citealt{Cornelisse2002}; \citealt{DelSanto2007}; \citealt{Chelovekov2007}; \citealt{Degenaar2009}).     \\

\swiftj is a new VFXT BH candidate (\citealt{Casares2011}, Corral-Santana et al. 2012 submitted, private communication) discovered with the Swift Burst Alert Telescope (BAT; \citealt{Barthelmy2005}) on 2011 January
28  \citep{Krimm2011}. In the following days it was observed with
several X-ray satellites as well as ground-based telescopes. The MPI/ESO
2.2m telescope at La Silla detected a new optical source within the
\textit{Swift}/BAT error circle. The magnitude difference compared to
archival Sloan Digital Sky Survey (SDSS) images (when the source was in quiescence) was $\sim$6~mag.
This indicates a Galactic origin, likely a low mass X-ray binary (LMXB) or a dwarf nova
outburst (Rau et al. 2011), although the X-ray spectrum pointed to a LMXB nature \citep{Krimm2011a}. 
The SDSS photometry indicated that the
quiescent counterpart was very red and that the companion was likely an M4 star. 
This resulted in a distance estimate of $\sim$1.5 kpc  \citep{Rau2011}.
In this work we present analysis of the \textit{Swift} X-ray and ultraviolet--optical (UV/optical) data of \swiftj along its 2011 outburst.

\section{Observations and analysis}

Immediately following its discovery \swiftj was monitored with the \textit{Swift} satellite \citep{Gehrels2004} throughout its outburst both with the X-ray telescope (XRT; \citealt{Burrows2005}) and the Ultraviolet/Optical telescope (UVOT; \citealt{Roming2005}), which lasted $\sim$7~months. A total of 42 pointings were performed during this period, for a total exposure time of 54.8~ks. A log of the observations is given in Table \ref{t1}.
The observations were performed daily in the first 10 days, and once every 2 days in the following month. After that, the observations were separated by 1 to 2 weeks.

\begin{table*}
\begin{center}
\caption[]{Log and XRT spectral results for \swiftj. }
\label{t1}
\begin{tabular}{ccccccccc}
\hline 
\hline
Obs-ID & Date & Mode & Exp-time (ksec) & $\Gamma$ & \Fx$_{,abs}\ ^{a}$ & \Fx$_{,unabs}\ ^{a}$ & $\lx\ ^{b}$& \chired /d.o.f. \\

\hline
31918002	&	2/1/11	&	WT	&	3.28	&	$	1.53	\pm	0.02	$	&	$	404	\pm	6	$	&	$	408	\pm	6	$	&	$	110	\pm	2	$	&	0.98	/	323	\\
31918003	&	2/2/11	&	WT	&	0.98	&	$	1.59	\pm	0.03	$	&	$	374	\pm	7	$	&	$	378	\pm	7	$	&	$	102	\pm	2	$	&	1.09	/	258	\\
31918004	&	2/3/11	&	WT	&	1.63	&	$	1.56	\pm	0.03	$	&	$	394	\pm	7	$	&	$	398	\pm	7	$	&	$	107	\pm	2	$	&	0.95	/	263	\\
31918005	&	2/4/11	&	WT	&	1.65	&	$	1.56	\pm	0.02	$	&	$	365	\pm	6	$	&	$	369	\pm	5	$	&	$	99.4	\pm	1.5	$	&	0.98	/	313	\\
31918007	&	2/6/11	&	WT	&	1	&	$	1.54	\pm	0.03	$	&	$	360	\pm	7	$	&	$	364	\pm	7	$	&	$	98.1	\pm	1.8	$	&	0.96	/	250	\\
31918008	&	2/7/11	&	WT	&	1.1	&	$	1.53	\pm	0.03	$	&	$	341	\pm	6	$	&	$	345	\pm	6	$	&	$	92.9	\pm	1.7	$	&	0.95	/	260	\\
31918009	&	2/8/11	&	WT	&	1.74	&	$	1.52	\pm	0.03	$	&	$	338	\pm	6	$	&	$	342	\pm	6	$	&	$	91.9	\pm	1.6	$	&	1.10	/	274	\\
31918010	&	2/9/11	&	WT	&	1.14	&	$	1.57	\pm	0.03	$	&	$	321	\pm	6	$	&	$	325	\pm	6	$	&	$	87.5	\pm	1.7	$	&	0.90	/	244	\\
31918011	&	2/10/11	&	WT	&	1.03	&	$	1.55	\pm	0.03	$	&	$	310	\pm	6	$	&	$	314	\pm	6	$	&	$	84.5	\pm	1.7	$	&	1.08	/	240	\\
31918012	&	2/11/11	&	WT	&	1.19	&	$	1.56	\pm	0.03	$	&	$	303	\pm	5	$	&	$	306	\pm	5	$	&	$	82.5	\pm	1.5	$	&	0.94	/	260	\\
31918013	&	2/16/11	&	WT	&	1	&	$	1.53	\pm	0.03	$	&	$	292	\pm	6	$	&	$	296	\pm	6	$	&	$	79.6	\pm	1.7	$	&	0.99	/	224	\\
31918014	&	2/18/11	&	WT	&	1.25	&	$	1.53	\pm	0.03	$	&	$	258	\pm	5	$	&	$	261	\pm	5	$	&	$	70.2	\pm	1.4	$	&	0.88	/	239	\\
31918015	&	2/20/11	&	WT	&	1.08	&	$	1.53	\pm	0.04	$	&	$	228	\pm	5	$	&	$	230	\pm	5	$	&	$	62.0	\pm	1.4	$	&	0.88	/	189	\\
31918016	&	2/24/11	&	WT	&	0.49	&	$	1.57	\pm	0.06	$	&	$	203	\pm	7	$	&	$	205	\pm	7	$	&	$	55.2	\pm	1.9	$	&	1.02	/	99	\\
31918017	&	2/26/11	&	WT	&	1.11	&	$	1.61	\pm	0.04	$	&	$	172	\pm	4	$	&	$	175	\pm	4	$	&	$	47.0	\pm	1.1	$	&	0.92	/	173	\\
31918018	&	2/28/11	&	WT	&	1.39	&	$	1.57	\pm	0.04	$	&	$	168	\pm	4	$	&	$	170	\pm	4	$	&	$	45.7	\pm	1.0	$	&	0.78	/	203	\\
31918019	&	3/2/11	&	WT	&	1.35	&	$	1.56	\pm	0.04	$	&	$	157	\pm	4	$	&	$	159	\pm	4	$	&	$	42.8	\pm	1.0	$	&	0.92	/	185	\\
31918020	&	3/4/11	&	WT	&	1.31	&	$	1.53	\pm	0.04	$	&	$	147	\pm	4	$	&	$	148	\pm	4	$	&	$	40.0	\pm	1.0	$	&	1.02	/	174	\\
31918021	&	3/6/11	&	WT	&	1.31	&	$	1.57	\pm	0.04	$	&	$	130	\pm	4	$	&	$	131	\pm	4	$	&	$	35.4	\pm	1.0	$	&	0.94	/	141	\\
31918022	&	3/8/11	&	WT	&	1.3	&	$	1.52	\pm	0.04	$	&	$	144	\pm	4	$	&	$	145	\pm	4	$	&	$	39.1	\pm	1.0	$	&	0.97	/	160	\\
31918023	&	3/10/11	&	WT	&	1.2	&	$	1.55	\pm	0.04	$	&	$	123	\pm	3	$	&	$	125	\pm	3	$	&	$	33.6	\pm	0.9	$	&	0.81	/	131	\\
31918024	&	3/12/11	&	WT	&	1.38	&	$	1.57	\pm	0.05	$	&	$	130	\pm	4	$	&	$	132	\pm	4	$	&	$	35.4	\pm	1.0	$	&	1.12	/	114	\\
31918025	&	3/13/11	&	WT	&	1.23	&	$	1.58	\pm	0.05	$	&	$	106	\pm	3	$	&	$	107	\pm	3	$	&	$	28.8	\pm	0.9	$	&	1.21	/	122	\\
31918026	&	3/16/11	&	WT	&	0.25	&	$	1.53	\pm	0.13	$	&	$	80.3	\pm	6.6	$	&	$	81.2	\pm	6.5	$	&	$	21.9	\pm	1.7	$	&	0.97	/	22	\\
31918027	&	3/18/11	&	WT	&	1.21	&	$	1.6	\pm	0.05	$	&	$	85.3	\pm	2.8	$	&	$	86.3	\pm	2.8	$	&	$	23.2	\pm	0.8	$	&	0.97	/	102	\\
31918028	&	3/20/11	&	WT	&	1.22	&	$	1.52	\pm	0.1	$	&	$	97.5	\pm	5.8	$	&	$	98.5	\pm	5.8	$	&	$	26.5	\pm	1.6	$	&	0.88	/	45	\\
31918029	&	3/29/11	&	WT	&	1.27	&	$	1.57	\pm	0.06	$	&	$	76.9	\pm	3.1	$	&	$	77.8	\pm	3.1	$	&	$	20.9	\pm	0.8	$	&	0.97	/	76	\\
31918030	&	4/3/11	&	WT	&	1.45	&	$	1.61	\pm	0.06	$	&	$	63.6	\pm	2.2	$	&	$	64.3	\pm	2.2	$	&	$	17.3	\pm	0.6	$	&	0.95	/	93	\\
31918031	&	4/13/11	&	WT	&	1.34	&	$	1.47	\pm	0.12	$	&	$	45.0	\pm	3.2	$	&	$	45.4	\pm	3.2	$	&	$	12.2	\pm	0.8	$	&	0.53	/	29	\\
31918032	&	4/20/11	&	WT	&	1.56	&	$	1.63	\pm	0.08	$	&	$	36.2	\pm	1.6	$	&	$	36.6	\pm	1.6	$	&	$	9.87	\pm	0.43	$	&	1.02	/	60	\\
31918033	&	4/23/11	&	WT	&	1.26	&	$	1.6	\pm	0.09	$	&	$	35.4	\pm	1.8	$	&	$	35.8	\pm	1.8	$	&	$	9.63	\pm	0.49	$	&	0.84	/	49	\\
31918034	&	4/28/11	&	WT	&	1.24	&	$	1.73	\pm	0.15	$	&	$	19.9	\pm	1.7	$	&	$	20.2	\pm	1.7	$	&	$	5.45	\pm	0.46	$	&	1.10	/	26	\\
31918035	&	5/3/11	&	WT	&	1.12	&	$	1.79	\pm	0.1	$	&	$	28.3	\pm	1.7	$	&	$	28.7	\pm	1.6	$	&	$	7.73	\pm	0.44	$	&	1.00	/	38	\\
31918036	&	5/8/11	&	WT	&	1.02	&	$	1.72	\pm	0.13	$	&	$	21.9	\pm	1.6	$	&	$	22.2	\pm	1.6	$	&	$	5.97	\pm	0.43	$	&	1.04	/	25	\\
31918037	&	5/13/11	&	WT	&	1.34	&	$	1.69	\pm	0.13	$	&	$	21.5	\pm	1.6	$	&	$	21.8	\pm	1.6	$	&	$	5.87	\pm	0.44	$	&	1.08	/	21	\\
31918038	&	6/15/11	&	PC	&	1.15	&	$	1.87	\pm	0.25	$	&	$	5.64	\pm	0.70	$	&	$	5.74	\pm	0.69	$	&	$	1.54	\pm	0.19	$	&	0.95	/	31	\\
31918039	&	6/29/11	&	PC	&	1.24	&	$	1.89	\pm	0.27	$	&	$	3.92	\pm	0.51	$	&	$	3.99	\pm	0.50	$	&	$	1.07	\pm	0.13	$	&	0.99	/	23	\\
31918040	&	7/14/11	&	PC	&	1.07	&	$	1.76	\pm	0.53	$	&	$	3.23	\pm	0.85	$	&	$	3.28	\pm	0.84	$	&	$	0.88	\pm	0.23	$	&	1.00	/	10	\\
31918041	&	7/28/11	&	PC	&	0.65	&	$	2.13	^{+ 1.86 }_{-0.95}	$	&	$	1.65	\pm	0.81	$	&	$	1.69	\pm	0.79	$	&	$	0.45	\pm	0.21	$	&	1.21	/	4	\\
31918043	&	8/23/11	&	PC	&	1.19	&		2.13 (fix)				&		$<$ 0.22				&		$<$ 0.23				&		$<$ 0.06				&		--		\\
31918044	&	9/5/11	&	PC	&	2.05	&		2.13 (fix)				&		$<$ 0.07				&		$<$ 0.08				&		$<$ 0.02				&	--			\\
\hline
\end{tabular}

\begin{list}{}{}
\item Note.- \Nh has been fixed to 1.2 $\times 10^{20}$ cm$^{-2}$, value inferred from the high-resolution X-ray spectra obtained with \textit{XMM-Newton} (Armas Padilla et al. 2012 in prep), which is consistent with that reported by \citet{Krimm2011a}.
\item $^{a}$ Flux in units of $10^{-12}$ erg cm$^{-2}$ s$^{-1}$ in the 0.5--10 keV energy band.
\item $^{b}$ X-ray luminosity in units of $10^{33}$ erg s$^{-1}$ calculated from the 0.5--10 keV unabsorbed flux by adopting a distance of 1.5 kpc.
\end{list}

\end{center}
\end{table*}

\subsection{XRT data}

The \textit{Swift}/XRT observations were obtained both in photon-counting (PC) and windowed timing (WT) modes. Due to the high flux of \swiftj the XRT was operated in WT mode in the early observations in order to avoid pile-up (except for the first observation; see below). When the count rate was below 0.5~count/s the observations were taken in PC mode (see Table \ref{t1}). 
The data were processed using the {\ttfamily HEASoft} v.6.11 software. The raw data were reduced running the {\ttfamily xrtpipeline} task in which standard event grades of 0--12 were selected for the PC data and 0--2 for the WT mode observations.\\

For every observation, spectra, lightcurves and images were obtained using {\ttfamily Xselect}. The regions to extract the source events both for PC and WT modes were centred at the source position (RA=13$^{h}$57$^{m}$16$.\!\!^{s}$86, Dec=-09\deg32$'$38\farcs9, J2000, 0\farcs42 error radius; \citealt{Krimm2011a}). For the WT data, we used a circle of $\sim$27~pixels of radius for the source and for the background extraction we used an annulus centred on the source with $\sim$64~pixels for the inner radius and $\sim$118~pixels for the outer radius. In the case of the PC mode observations, the region used was a circle of $\sim$10~pixels of radius for the source and for the background three circular regions with the same size located in nearby source-free regions.\\ 

To look for eclipses, dips or X-ray bursts we checked the lightcurves with different bin--sizes extracted for each observation. We did not detect any evidence or feature that suggest modulations or a pattern in the source flux, although we cannot rule out with high confidence the presence of any modulations due to the low count rate.
We created the exposure maps in order to correct for the effect of bad columns on the CCD, which in turn are used to create the ancillary response files using the {\tt xrtmkarf} task. The latest versions of the response matrix files (v.14) were taken from the HEASARC calibration database. Finally, with {\tt grppha} we grouped the spectra to have a minimum of 20 photons per bin to be able to consistently use the \chis. Due to the low number of counts collected during observations 38/39/40/41\footnote{We refer to the observations with the last two numbers of their observation-ID's in the main text. The full observation--ID's are in Table \ref{t1}}, we fitted these data using both  C-statistic and \chis but grouped them with a minimum of 5 photons per bin. The results using both methods were consistent with each other.\\

The first observation 01 was taken in PC mode which is severely affected by pile-up. It has a count rate of 2.3~c/s, while the pile-up becomes considerable in PC mode for count rates above $\sim$0.6~c/s  (\citealt{Evans2009}; see also the \textit{Swift}/XRT analysis web\footnote{http://www.swift.ac.uk/analysis/xrt/xselect.php}). Therefore we do not include this observation in our analysis. In the last two XRT observations (43/44) the source was not detected. The count rates were calculated using the prescription for small numbers of counts given by \citet{Gehrels1986}. We calculated the 95 per cent confidence upper limits on the flux with WebPIMMS HEASARC tool\footnote{Available from http://heasarc.gsfc.nasa.gov/Tools/w3pimms.html}. An absorbed power-law model with a hydrogen column density (\Nh) of 1.2$\times 10^{20}$ cm$^{-2}$ (see Section 3) was assumed. We used a photon index of $\Gamma$=2.13 which corresponds to the value we obtained in the last observation where the source was detected.\\

In order to improve the statistics we have also combined observations that were performed within a few days time span and that yielded comparable fluxes. As such, we have combined observations 03--05, 07--12, 13--15, 17--20, 21--25, 26--20, 30--33, 34--37, 38--41 and 43--44. We extracted the combined  spectra, lightcurves and images in the same way as the individual cases. We merged the individual exposure maps created for each image using the task {\tt ximage}.

\subsection{UVOT data}

The UVOT observations were performed in image mode. Most of them were taken with 6 filters (\textit{v, b, u, uvw1, uvm2, uvw2}) but some of them with only a few filters and sometimes only one. We calculated the source's magnitude (in the Vega system) and flux densities with the {\tt uvotsource} tool, which performs aperture photometry on the sky images. We selected a circular region with a radius of 5~arcsec centred on the source and a circular region source--free with a radius of 18~arcsec for the background correction. We corrected the magnitudes and fluxes for the Galactic extinction. The reddening is \textit{E(B--V)}=0.04~mag in the direction of \swiftj \citep{Schlegel1998}. Using this value and the prescription of \citet{Pei1992} we calculate the extinction for every band. The obtained values are \textit{A$_{v}$}=0.123, \textit{A$_{b}$}=0.163, \textit{A$_{u}$}=0.193, \textit{A$_{uvw1}$}=0.263, \textit{A$_{uvm2}$}=0.387 and \textit{A$_{uvw2}$}=0.349. The source is located at high Galactic latitude (b=50.0042) and for a distance of 1.5 kpc or larger, the source will be located outside the plane of the Galaxy. Therefore, the measured column density (or reddening) corresponds to the full Galactic one in that direction. In case of no source detection the {\tt uvotsource} tool returns a 3--sigma upper limit for the magnitudes.

\section{Results}

\begin{figure*}
\begin{center}

\includegraphics[angle=0,width=\columnwidth]{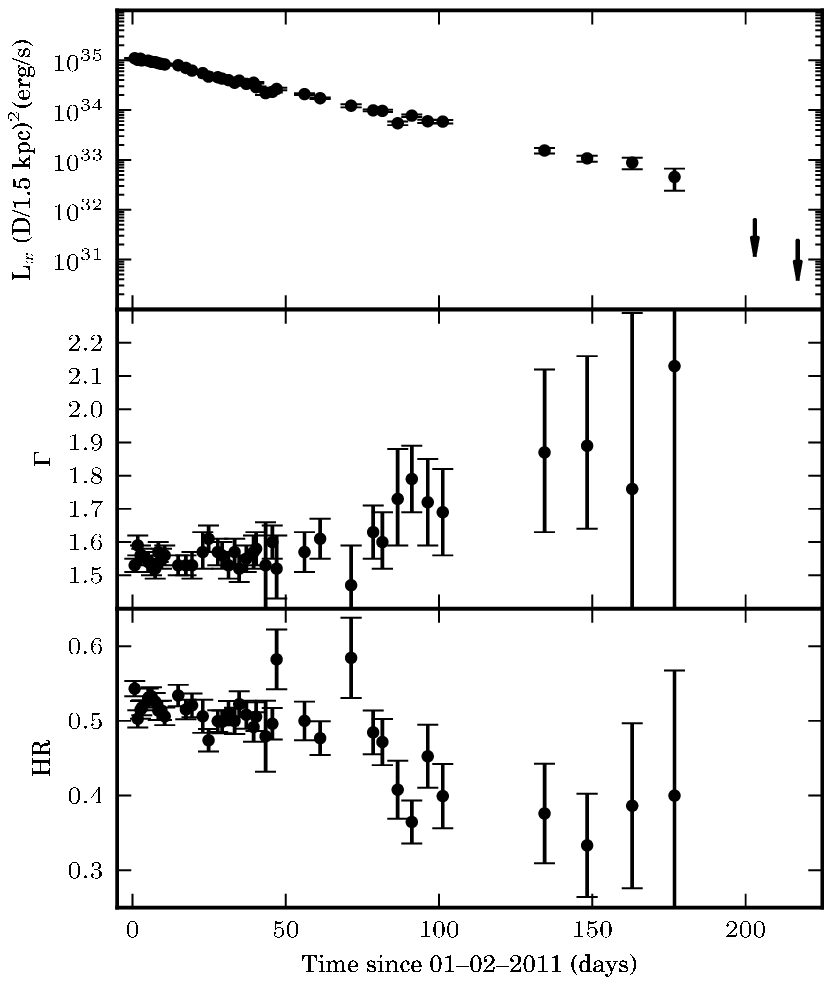}
\includegraphics[angle=0,width=\columnwidth]{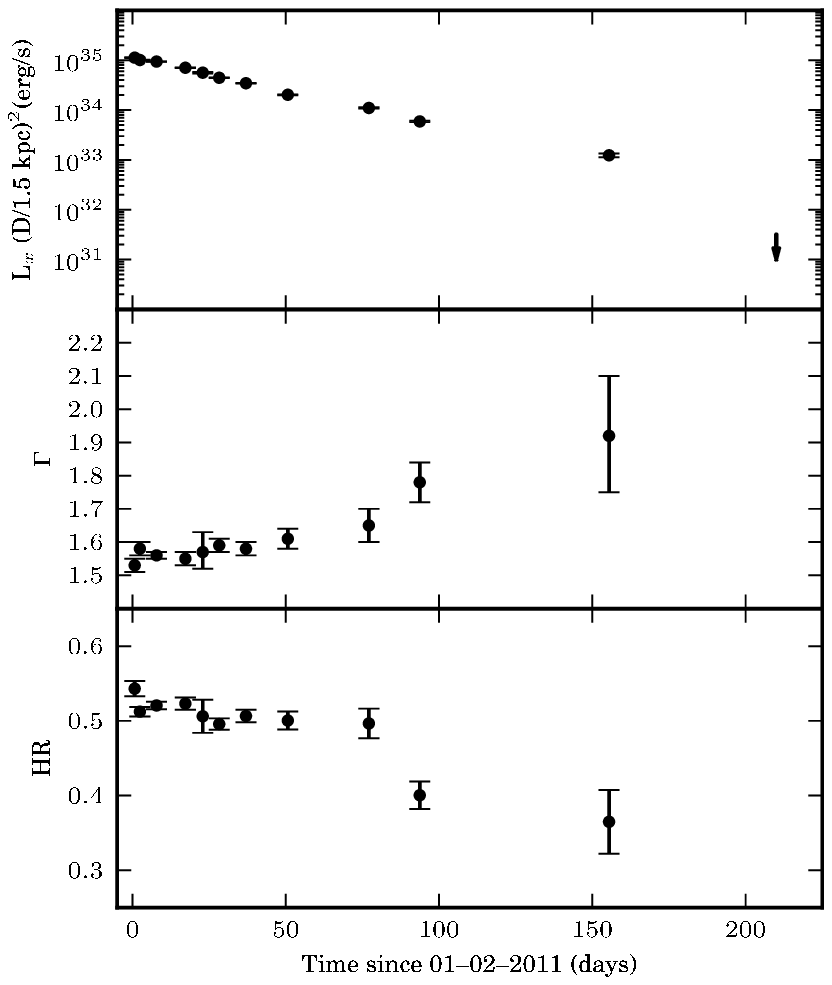}

\caption{In both plots the upper panel is the light curve (0.5-10 keV; assumed distance of 1.5~kpc), the central panel is the photon index evolution with time, and the bottom panel is the  hardness ratio evolution with time (ratio of the counts in the hard, 2-10 keV, and soft, 0.5-2 keV energy bands). All data are plotted in the left figure, whereas in the right figure we present the result of the combined sets of observations (see Section 2.1). }
\label{f1}
\end{center}
\end{figure*}

\subsection{X-ray light curve and spectra}
We fitted the spectra using {\ttfamily XSPEC} (v~12.7.0). All observations were well fitted with a simple power--law ({\tt powerlaw}) affected by photoelectric absorption ({\tt phabs}). We assumed a column density to \swiftj of 1.2~$\times 10^{20}$cm$^{-2}$ constant with time.  We inferred this value from the high-resolution X-ray spectra obtained with \textit{XMM-Newton} (Armas Padilla et al. 2012 in prep), which is consistent with that reported by \citet{Krimm2011a}. The results are reported in Table \ref{t1}, where the flux errors have been calculated following the procedure presented by \citet{Wijnands2004}. We also tried other single component models like a blackbody ({\tt bbodyrad}) and an accretion disk consisting of multiple blackbody components ({\tt diskbb}, \citealt{Makishima1986}) but both models could not fit the data accurately (\chired$>$ 3 and \chired$>$ 1.5, respectively). A two component model ({\tt diskbb+powerlaw}) can fit the data as well, however it does not improve the best fit of the single power--law model and the additional thermal component contributes only a few per cent of the total flux.\\

In Fig. \ref{f1} we show the 0.5--10 keV light curve. The outburst has its maximum as observed during observation 02 in the beginning\footnote{Observation 01 could be brighter, but due to the strong pile-up it was not included in the analysis.}, where the peak of the unabsorbed flux is  4.1~$\times$~$10^{-10}$~erg~cm$^{-2}$s$^{-1}$. This corresponds to a peak luminosity of 1.1~$\times$~$10^{35}$~erg~s$^{-1}$ assuming a distance of 1.5~kpc \citep{Rau2011}. After this, the luminosity monotonically decreases until the source goes undetected using the XRT $\sim$180 days after the discovery date. The upper--limits on the luminosity calculated from the last two observations are between 2 and 6~$\times$~$10^{31}$~erg~s$^{-1}$.\\

Plotting the photon index evolution with time (Fig. \ref{f1}) shows that the photon index increases from a value of $\Gamma\sim$1.5 to a value of $\Gamma\sim$2 indicating that the spectra became softer during the decay of the outburst.
This softening behaviour is also seen when using the hardness ratio (HR; Fig.\ref{f1} left panel). The hardness ratio is defined as the ratio between the counts in the hard band (2--10 keV) and the counts in the soft band (0.5--2 keV).
We have performed the same analysis when combining sets of observations (see Section 2.1). The result is shown in Fig. \ref{f1} right panel. Due to the smaller error bars, the increase of the photon index and the HR decay are more clearly visible.\\

\subsection{Ultraviolet/optical and X-ray correlation}

\begin{figure}
\begin{center}

\includegraphics[angle=0,width=\columnwidth]{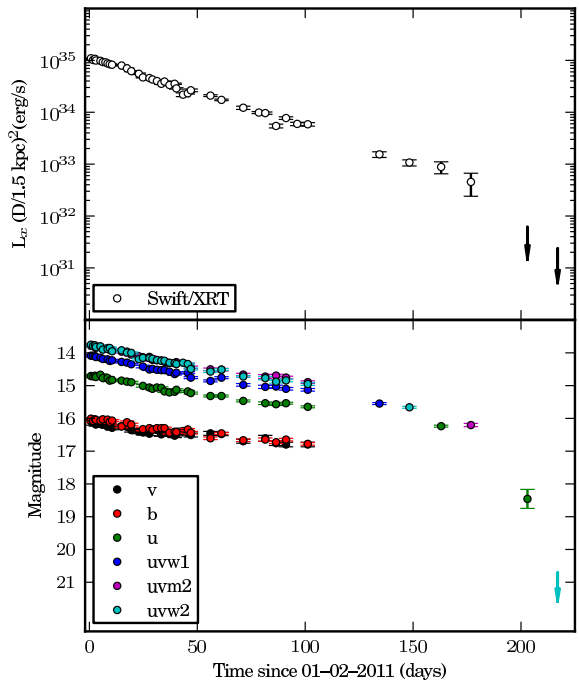}

\caption{ In the top panel the X-ray light curve in the 0.5--10 keV energy band for a distance of 1.5~kpc is shown while in the bottom panel the UV/optical light curves in the Vega system are plotted}
\label{f2}
\end{center}
\end{figure}

\begin{figure}
\begin{center}
\includegraphics[angle=0,width=\columnwidth]{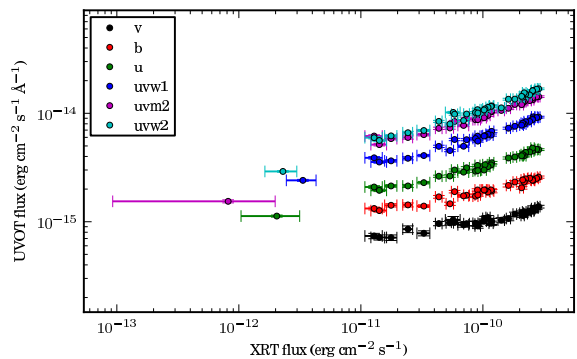}
\caption{Correlation between the UV/optical density fluxes and the X-ray 2--10~keV flux.}
\label{f3}
\end{center}
\end{figure}

\begin{figure}
\begin{center}
\includegraphics[angle=0,width=\columnwidth]{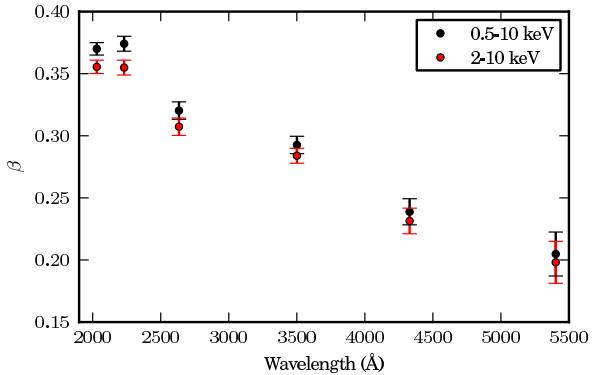}
\caption{Correlation slopes ($\beta$) between the UVOT bands and the X-ray flux in the 0.5--10~keV and 2--10~keV energy bands.}
\label{f4}
\end{center}
\end{figure}

The \textit{Swift}/UVOT observations were taken simultaneously with the X-ray ones. Most of the  observations were taken with all six filters, which allows us to study the correlation between the X-ray and the UV/optical emission along the outburst. Fig. \ref{f2} shows the X-ray light curve and the UV/optical magnitudes in the Vega system. The last UVOT detection (\textit{u} band) was 203~days after the first detection when the source was already undetected in X-ray. Over the course of the outburst the brightness in all bands is decreasing simultaneously with the decline in the X-rays. This resulted in a clear correlation between X-ray and UV/optical fluxes which is shown in Fig. \ref{f3}. We have assumed a power-law model to fit these correlations and calculate the correlation slopes $\beta$ (F$_{UV/optical}\propto$~F$_{X}$~$^{\beta}$).
We have calculated $\beta$ for every UVOT band and for the X-ray flux in the 2--10 keV energy range in order to compare them with values in previous publications (e.g. \citealt{Russell2006}). However, the spectral resolution of \textit{Swift}/XRT allows us to also measure the correlation slopes for X-ray fluxes in the 0.5--10 keV energy range. The results are shown in Table \ref{t3}. Clearly $\beta$ increases towards shorter wavelengths, going from 0.2 in the \textit{v} band to 0.37 \textit{uvw2} (using the 0.5--10~keV energy range). The numbers become 0.19 to 0.35 in the 2--10~keV range (see Fig. \ref{f4}).   

\begin{table}
\begin{center}
\caption[]{Correlation slope between UV/optical and X-ray fluxes (F$_{UV/optical}\propto$~F$_{X}$~$^{\beta}$).  $\lambda_{eff}$ is the effective wavelength for each band.}
\label{t3}
\begin{tabular}{cccc}
\hline 
\hline
UVOT/Band& $\lambda_{eff}$ &\multicolumn{2}{c}{$\beta$} \\
\cline{3-4}
&(\AA) & 0.5--10 KeV & 2--10 keV \\
\hline  
v & 5402 &$0.203 \pm	0.019$ &	$0.199 \pm	0.017$ \\
b &	4329 &$0.237	\pm 0.011$ &	$0.232 \pm	0.011 $\\
u &	3501 &$0.293 \pm	0.007$ &	$0.286 \pm	0.006$ \\
uvw1 & 	2634 &$0.321 \pm	0.007$ &	$0.311 \pm	0.006$ \\
uvm2 & 	2231 &$0.374 \pm	0.007$ &	$0.357 \pm	0.006$ \\
uvw2 & 	2030 &$0.370 \pm 0.006$ &	$0.358 \pm	0.005$ \\
 
\hline
 
\end{tabular}
\end{center}
\end{table}

\section{Discussion}

We have analysed the XRT and UVOT data taken over the $\sim$7~months duration of the 2011 outburst of the newly discovered candidate black hole LMXB \swiftj. We fitted the \textit{Swift}/XRT spectra using a simple power-law model affected by absorption. The peak flux was reached at the beginning of the outburst, after which, it steadily decreased until the source went undetected with XRT $\sim$180 days after its discovery. The unabsorbed peak flux has a value of \Fx$_{,unabs}$ = 4.1$\times$ $10^{-10}$~erg~cm$^{-2}$s$^{-1}$ in the 0.5--10 keV energy range, which corresponds to an outburst peak luminosity of $\lx$=1.1$\times$ $10^{35}$~erg~s$^{-1}$ assuming a distance of 1.5~kpc. This low peak luminosity classifies \swiftj as a VFXT, which have peak luminosities $\lx^{peak}<$10$^{36}$~erg~s$^{-1}$. Even if the source was located at a distance of 8~kpc the inferred peak luminosity is a few times 10$^{36}$~erg~s$^{-1}$, still in the faint regime. However, such long distance is unlikely since the high Galactic latitude (b=50.0042) would place \swiftj at 6 kpc above the Galactic plane.\\

The system was in the hard state at peak of the outburst and remained in this state throughout the outburst, which is usual at these low X-ray luminosities (see \citealt{Belloni2010} for the state definitions). From the first observation until the last one the spectral index of the power law increases from 1.5 to 2.1. As can be seen in Fig. \ref{f1}, the X-ray luminosity and photon index are anti-correlated. The softening can also be seen in the HR curve demonstrating that it does not depend on assumed spectral model. It shows how the spectrum becomes softer as the luminosity decays (Fig. \ref{f1}).
Such softening behaviour was also present in the VFXT XTE~J1719--291 \citep{ArmasPadilla2011}, but for this source the spectra were considerably softer, with a photon index regime from $\Gamma$=2.0 to $\Gamma$=2.7. The reason for \swiftj to have harder spectra than XTE~J1719--291 could be a difference in the nature of their compact objects. Although it is not established, the characteristics of XTE~J1719--291 favour a NS accretor \citep{ArmasPadilla2011} whereas \swiftj has been identify as a strong BH candidate system using optical spectroscopy (\citealt{Casares2011}, Corral-Santana et al. 2012 submitted, private communication).\\

The upper-limit on the luminosity inferred from the non-detections in the last two observations is 
$L_{X}$~$\sim$~2$\times$10$^{31}$~erg~s$^{-1}$. This could indicate that the nature of the accretor is a BH since they typically have quiescent luminosities in the range of $\lx\sim10^{30}-10^{31}$~erg~s$^{-1}$ while the NS's usually have quiescent luminosities $\lx>$10$^{32}$~erg~s$^{-1}$ (but see \citealt{Jonker2007} for an exception).\\

Several other BH sources have shown softening in their X-ray spectra at low luminosities. XTE~J1650--500 softened in its 2001-2002 outburst. The photon index changed from $\Gamma$=1.66 to $\Gamma$=1.93 when the luminosity decayed to 2$\times$10$^{34}$~erg~s$^{-1}$ \citep{Tomsick2004}. Also at these intermediate luminosities 4U~1543--47 evolved from a photon index of $\Gamma$=1.64 in its hard state to a photon index of $\Gamma$=2.22 at lower $L_{X}$ \citep{Kalemci2005}. By studying a sample of 7 BHs in their quiescence state, \citet{Corbel2006} arrived to the conclusion that all BHs are softer in quiescence than in the standard hard state. Using a study of 25 BHs, \citet{Dunn2010} presented data from which the same conclusion can be inferred: when data are available, the sources show softening towards quiescence. However, in the 2008 outburst decay of H~1743--322 to quiescence, \citet{Jonker2010} did not see any evidence of softening, although it cannot be ruled out that this is caused the inaccuracy in the photon index due to the high \Nh.

\subsection{X-ray and UV/optical simultaneous data: A non-irradiated accretion disc}

The proximity and the low Galactic extinction towards \swiftj have made it possible to obtain X-ray and UV/optical data simultaneously using XRT and UVOT instruments on board \textit{Swift}. Until today this had not been possible for any VFXT since, in addition to their low luminosity, the vast majority are located towards the Galactic centre, which makes it very difficult to detect the optical counterpart due to the high Galactic absorption.
\\
The optical/UV and X-ray fluxes are strongly correlated during the outburst (Fig. \ref{f2}). \citet{Russell2006} quantified the disc and jet contribution from the optical/infrared and X-ray (2-10 keV) correlation in the hard state. Based on the values of $\beta$ expected from the different emission processes (sections 3.3 and 3.4 of their work), our correlation in the \textit{v} band is consistent with a BH accreting via a non irradiated viscous disc. For a viscously heated disc, $0.15 \leq \beta \leq 0.25$ is expected in the case of a BH (for \textit{V}-band optical and 2--10 keV X-ray), and for \swiftj we find $\beta = 0.2$ (Table 2). For a NS, $0.3 \leq \beta \leq 0.5$ is predicted, so our correlation is consistent with optical emission from a viscously heated disc around a BH accretor. Higher values of $\beta$ are expected if the optical emission originates in an irradiated accretion disc or a jet.
\\
Additionally, \citet{Frank2002} predict a dependency of the slopes $\beta$ with the wavelength. For optical emission from a viscously heated steady-state disc, $\beta$ increases at shorter wavelengths. This is consistent with our results, where $\beta$ increases from 0.2 in the \textit{v} band to 0.35 in the \textit{uvw2} one (see Table \ref{t3}). Therefore, all the evidence favours UV/optical emission dominated by the viscously heated disc, with little or no X-ray irradiation detected.\\

\section*{Acknowledgments}

MAP acknowledges J. Corral-Santanta and J. Casares for useful discussion on the nature of the source. MAP and RW are supported by an European Research Council Starting Grant awarded by RW. RW also acknowledge support from The European Community's Seventh Framework Programme (FP7/2007-2013) under grant agreement number ITN 215212 Black Hole Universe. ND is supported by NASA through Hubble Postdoctoral Fellowship grant number HST-HF-51287.01-A from the Space Telescope Science Institute (STScI). D.M.R is supported by a Marie Curie Intra European Fellowship within the 7th European Community Framework Programme under contract No. IEF 274805.

\bibliographystyle{mn2e}
\bibliography{bibliography}

\end{document}